\documentclass[]{article}
\usepackage[a4paper, total={6in, 8in}]{geometry}
\usepackage{authblk}
\usepackage{natbib}
\usepackage{graphicx}
 \usepackage{amsmath}

\title{The 24 Aqr triple system: A closer look at its unique high-eccentricity hierarchical architecture}

\author{Ahmad Abushattal$^1$, Mashhoor A. Al-Wardat$^{2,3}$, Elliott P. Horch$^4$, Nikolaos Georgakarakos$^{5,6}$, Hatem A. Al-Ameryeen$^1$, Enas M. Abu-Alrob$^3$, \\and Abdallah M. Hussein$^3$}
\affil[1]{Department of Physics, Al-Hussein Bin Talal University, PO Box 20, 71111, Maan, Jordan}
\affil[2]{Department of Applied Physics and Astronomy, and Sharjah Academy for Astronomy, Space Sciences and Technology, University of Sharjah, PO Box 27272, Sharjah, UAE}
\affil[3]{Department of Physics, Al al-Bayt University, Mafraq 25113, Jordan}
\affil[4]{Department of Physics, Southern Connecticut State University, 501 Crescent Street, New Haven, CT 06515, USA}
\affil[5]{Division of Science, New York University Abu Dhabi, Abu Dhabi, UAE}
\affil[6]{Center for Astrophysics and Space Science (CASS), New York University, Abu Dhabi, PO Box 129188, Abu Dhabi, UAE}

\begin{document}

\maketitle

\begin{abstract}
As its periastron passage occurred during the third quarter of 2020, system 24 Aqr is of particular signiﬁcance. New visual solutions
for the latest speckle interferometry observations collected by the Lowell Discovery Telescope (LTD) with its new QWSSI speckle camera
are presented here. A variety of techniques were used to analyze the system, including ORBITX code for orbital solution, Al-Wardat’s
method for analyzing multiple stellar systems, and Edwards’ method for analyzing visual and spectroscopic binaries. We derive precise
masses and the complete set of its fundamental parameters for the three components, and we introduce a new orbital solution, and a new
dynamical parallax, which is very close to the measured value given by Hipparcos 2007 and from that of Gaia DR2. In the next section,
we discuss the possibility of a coplanar orbit. In conclusion, we demonstrate that we need a 65-m telescope to resolve the inner binary
visually, although an array of telescopes could be used instead.
\end{abstract}

\section{Introduction}

The study of multiple stellar systems has gained special
importance over time because of its inﬂuence on understanding 
the stellar formation and evolution theories. The rapid development 
of the observational tools and the availability of data provided 
by astronomical missions such as
Gaia, Hipparcos, and Kepler, helped to reveal the proper-
ties of such systems (Perryman et al., 1997; Van Leeuwen,
1997; Prusti et al., 2016). The prime objective of the Kepler
mission was to search for Earth-like planets, but it also
detected eclipses in multiple systems (Carter et al., 2011;
Abushattal et al., 2019; Abushattal et al., 2022a;
Abushattal et al., 2022b). More than $50\%$ of the stars in
the galaxy are believed to be binary or multiple systems,
with $46\%$ of them having at least one companion, and more
than $30\%$ are multiple stellar systems with a binary star as
one of the components (Stassun, 2012; Tokovinin, 2014b).
During the past decades, many studies have used high-
resolution techniques to investigate those binaries. Those
eﬀorts provided the most appropriate methods for calculating 
and determining orbital parallaxes and stellar masses
by using the correlation between the visual and spectroscopic 
orbits (Bonneau et al., 1980; Balega and
Ryadchenko, 1984; Fekel et al., 1997; Al-Wardat et al.,
2017; Docobo et al., 2017; Mendez et al., 2018;
Tokovinin, 2018; Docobo et al., 2018a; Docobo et al.,
2018b).

Many multiple systems are hierarchical triples (Abu-Alrob et al., 2023). 
They can be perceived as being comprised of two subsystems: a single 
star (the outer star) that
orbits widely around the center of mass of a binary system
(inner binary) with $a_{out}>>a_{in}$, $a_{out}$ and $a_{in}$ being the semi-
major axis of the outer and inner orbit respectively. The
evolution of such systems is governed by various astrophysical 
processes taking place among its members, such
as for example mass transfer, tidal friction, and gravitational 
interactions (e.g. see Georgakarakos, 2002; Fabrycky and Tremaine, 2007; Toonen et al., 2020; Naoz et al., 2013; Taani et al., 2019a; Taani et al., 2019b). Triple
systems can help us understand star formation and evolution during the various phases. Fekel (1981), Vynatheya et al. (2022), Georgakarakos (2005) published the ﬁrst list
of hierarchical systems. Since then, several observational
eﬀorts have been made to study that kind of system (e.g.
Fekel et al., 1997; Derekas et al., 2011; Tokovinin, 2014b;
Tokovinin, 2014a).

The system 24 Aqr consists of a close spectroscopic binary (SB) and a third star on a much wider orbit. 24 Aqr was known to be a triple as early as 1963 (Eggen, 1963), where it was also suggested that the system’s total mass was $1.51 M_{\odot} $. Based on the variability of the system’s radial velocity measurements, the same conclusion regarding the triple nature of the system was reached by Heintz (1981). Later on, Lippincott (1982) suggested a 
$0.73\pm0.20 M_{\odot}$ star for the primary component, while the secondary component
was $0.76\pm24 M_{\odot}$. Based on the amplitude of the radial
velocity curve, Griffin et al. (1996) determined that the
masses of the two components of the spectroscopic pair
were $1.19 M_{\odot}$ and $0.15 M_{\odot}$ , while the mass of the distant
visual component was $1.10 M_{\odot}$. The spectral type of the
system has been another subject of investigation. In the
Henry Draper Catalogue, the spectral type is F8V
(Cannon and Pickering, 1924), while it is F7V in Christy
and Walker (1969), and F8V in Harlan (1974). In SIMBAD, 
the F6V classiﬁcation was determined by Abt (1985), Wenger et al. (2000). 
Griffin et al. (1996) also studied this system and by combining the relative dip area of
the radial velocity based on spectroscopic observation
and the color index $B-V$, they suggested that the spectral
type for 24 Aqr was F8V, with F7V for the primary component of the spectroscopic binary and F9.5 V for the secondary.

The orbit of the visual binary has also been under investigation for over 100 years. Initially, Kuiper (1926) determined the orbital period to be 71 years with aneccentricity of 0.893. Subsequently, Finsen (1929b) gave a 51.33-year period with a 0.910 eccentricity, and then
Aitken (1932) brought the period down to 46.6 years and
the eccentricity to 0.85. A few years later, Mannino
(1946) proposed a much higher value of 99.46 years but
with an eccentricity of 0.25. In more recent years, Griffin
et al. (1996) found an orbit with a period of 48.7 years
and an eccentricity of 0.86. Finally, in 2019, Tokovinin
published the orbital solution in double stars information
circular No. 199 with a period of 48.67 years, based on
observations of (Aitken and Millard, 1932; Finsen,
1929a; Danjon, 1942; Mannino and Humblet, 1955;
Heintz, 1997; Scardia et al., 2019). Table 2 lists the orbital
elements for the visual orbital solutions mentioned above.

In this work, we reanalyze the system 24 Aqr, also
known as HD 206058 using the latest available data. We
attempt to determine as accurately as possible the physical
and orbital parameters of the 24 Aqr triple system. We
make use of all available data, spectroscopic and astrometric observations, and derived orbital solutions of the system. We shall apply two diﬀerent methods. First, we use
Edwards’ method in order to determine the individual
masses and spectral type of both components of the spectroscopic binary (Edwards, 1976; Abushattal et al., 2020). Subsequently, we use the Al-Wardat method which is an
atmospheric modeling and synthetic photometry method
with the aim of determining the physical parameters of
the system (Al-Wardat, 2003b; Al-Wardat, 2007; Al-
Wardat, 2008; Al-Wardat and Widyan, 2009; Al-Wardat
et al., 2014; Masda et al., 2019; Al-Wardat et al., 2021b;
Hussein et al., 2022).

Then, by combining the results of both methods, we get
the complete solution for the 24 Aqr triple system and we
are able to determine the most probable 3D orbit for the
system. Furthermore, we discuss the possibility of coplanarity of the system using all information at our disposal
from both the spectroscopic and visual binary orbits. The
structure of the paper is as follows: in Section 2 we analyze
the main component (A) of this triple system as a spectroscopic binary using Edwards' process and atmospheric modeling and synthetic photometry. In Section 3 we discuss the orbital alignment of the two orbital planes, while
in Section 4 we investigate whether we can resolve the spectroscopic binary component (A) visually. Finally, in Section 5, we present our conclusions.

\section{Analysis of system}

\subsection{The visual orbit}

The orbital properties for the main system of 24 Aqr
were determined using the dynamical technique developed by Tokovinin et al. (2016). This method uses the least-squares ﬁts with weights inversely proportional to
the observational errors to give the ﬁnal orbital parameters with their errors using ORBITX code with IDL. We used relative position measurements from speckle
interferometric observations to determine the visual
orbit. We obtained most of the positional measurements
for the system from the Fourth Catalog of Interferometric Measurements of Binary Stars (Hartkopf et al., 2010). However, four new measurements, used for the
ﬁrst time in this work, were taken with the QWSSI
speckle instrument (Clark et al., 2020) at the Lowell Discovery Telescope on 27 Aug 2021 UT. These were especially important to the visual orbit calculation as they
were close to the periastron passage. The full list of measures used is shown in Table 1. Fig. 1 shows the modiﬁed orbit of 24 Aqr. We used seven new measurements to
improve the orbital solution of the main orbit of the system. The new measurements covered the periastron passage which raised the orbit Grade from 2 to 1. A zero
residual signiﬁes an exact match with the orbit. We
therefore assign zero uncertainty to separation uncertainties for certain data points, while uncertainty for
position angle is not taken into account when calculating
orbital elements.

\subsubsection{The Spectroscopic Binary Component (A)}
In this section, we determine the three-dimensional orbit
for the SB components and we estimate the most probable
values of the physical parameters of this system such as the
magnitude, spectral type, and masses. For the single-lined
spectroscopic binary, we know the following orbital
parameters: the period P, the epoch of the periastron T,
the eccentricity e, A1= a1 sin(iS ) (a1 is the semi-major axis
of the orbit for the main component), the inclination iS ,
the argument of periastron x corresponding to the orbit
of the main component and the mass function $f(m)$. In
addition, we know the composite spectrum, the apparent
global magnitude, and the parallax (Hipparcos or Gaia
parallax). The initial data for 24 Aqr can be found in
Table 3.

Using Edwards’ method, we start from the transformation of the apparent magnitude to the absolute magnitude and then we apply Edward’s process in order to determine
the individual spectrum for each star depending on the difference of the magnitude Dm (Edwards, 1976; Abushattal et al., 2020; Docobo et al., 2017). We can also determine
the masses of each component, since, due to the improvement in observational methods in the last decades, the accuracy in the masses has increased signiﬁcantly. Hence,
we estimate the masses depending on the diﬀerent calibrations done by (Straižys and Kuriliene, 1981; Malkov, 2007;Docobo and Andrade, 2006; Pecaut and Mamajek, 2013).
Afterwards, we calculate the semi-major axis of the relative
orbit a and the two semi-major axes of the barycentric
orbits a1 and a2 in (A.U). Finally, from the semi-major
axes, and using the mass function, we can determine the
inclination, $i_S$ .

\subsection{Atmospheric Modeling and Synthetic Photometry}

One of the most eﬀective methods for the analysis of
such systems is the novel computational spectrophotometric method known as Al-Wardats’ method for analyzing
multiple systems (MSs) (Al-Wardat, 2002; Al-Wardat,
2012), which was successfully applied in the analysis of several main-sequence and subgiant BMSs (Al-Wardat,
2003b; Al-Wardat, 2003a; Al-Wardat, 2007; Al-Wardat,
2012; Al-Wardat et al., 2014; Masda et al., 2019; Al-
Wardat et al., 2021a; Hussein et al., 2022; Tanineah
et al., 2023).
The idea of the method in brief is to use the measured
visual magnitude mv and a color index $B-V$ of the entire
system along with the visual magnitude difference between
their components to estimate the complete set of their
physical and atmospheric parameters.
It is diﬃcult to measure the spectra for each component
of MSs, where the components are close to each other and
exhibited as one star in large telescopes. In order to get the
entire synthetic SED of the system which is related to the
energy ﬂux of the components located at a distance $d(pc)$
from the Earth, we use the following equation Al-Wardat
(2002):
\begin{eqnarray}
F_\lambda \cdot d^2 = (H_\lambda ^{Aa} \cdot R_{Aa} ^2+H_\lambda ^{Ab} \cdot R_{Ab} ^2)+ H_\lambda ^B
\cdot R_{B} ^2 
\end{eqnarray}
where $F_\lambda$ is the flux for the entire synthetic SED  of the entire binary system at the Earth, $H_\lambda ^Aa $, $H_\lambda ^Ab $ and  $H_\lambda ^B$ are the fluxes of the primary and secondary  components, while $ R_{Aa}$, $ R_{Ab}$ and $ R_{B}$ are the radii of the primary and secondary components in solar units $R_{\odot}$.

It employs Kurucz (ATLAS9) line-blanketed plane-
parallel model atmospheres to build the ﬂuxes for the individual components (Kurucz, 1994d) and atomic data provided later on (Kurucz, 1994a; Kurucz, 1994c; Kurucz,
1994b; Kurucz, 1994e). In order to apply the method, we need the magnitude diﬀerence between the components of the system, which was measured for the main two components A and B using speckle interferometry, and for the two sub-components Aa and Ab through the spectroscopic
analysis in this work as Edwards’ process from Table 6.
The synthetic spectral energy distributions (SEDs) for the
entire and individual components are shown in Fig. 2 (left one). Moreover, the results of synthetic photometry for Johnson-Cousins ($UBVR_C$) ﬁlters, Strömgren (ubvy) ﬁlters
and Tycho ﬁlters ($B_T, V_T$) for synthetic SEDs and color
indices are listed in Table 4. The atmospheric parameters
and other calculated fundamental parameters are listed in
Table 5. Fig. 2 shows the positions of the components on
the evolutionary tracks and isochrones Girardi et al.
(2000).

Where the metalicity Z can be calculated using measurements of Iron abundance relative to the Sun $[Fe/H]$ using the following equation:

\begin{equation}
\log{Z}=0.977[Fe/H]-1.699
\end{equation}

\begin{figure}
\centering
\includegraphics[width=150mm,height=160mm]{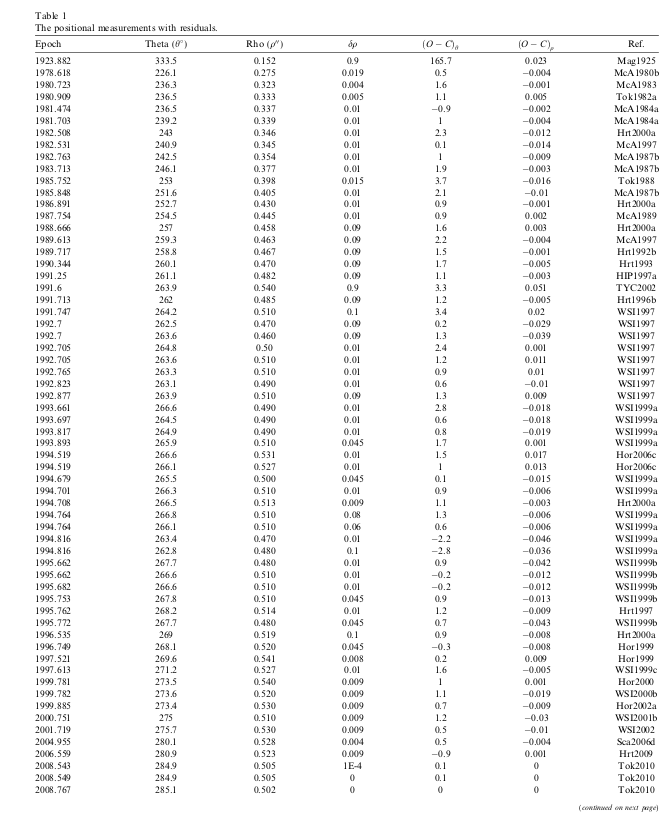}
\end{figure}

\begin{figure}
\centering
\includegraphics[width=145mm,height=40mm]{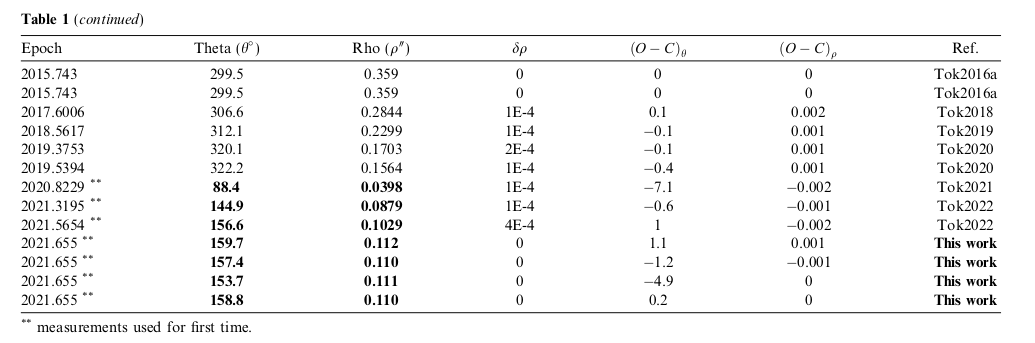}
\end{figure}

\begin{figure}
\centering
\includegraphics[width=100mm,height=100mm]{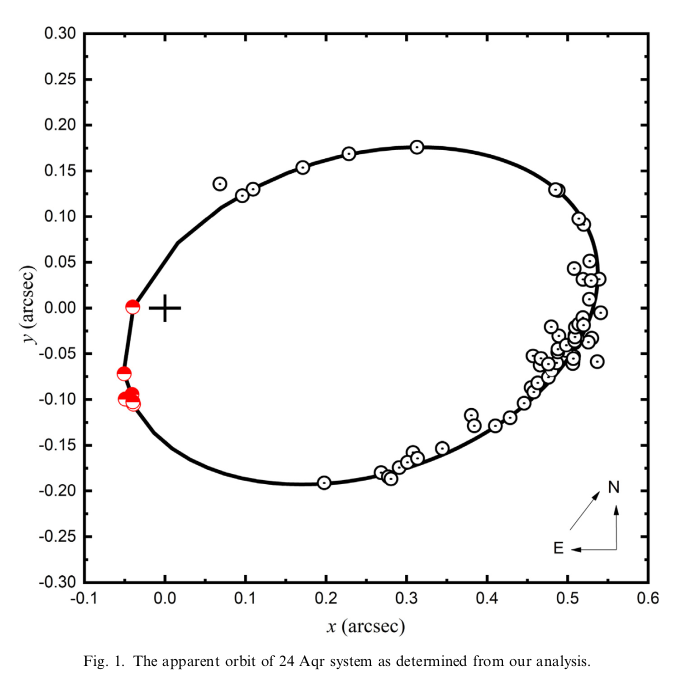}
\end{figure}

According to Gaspar et al. (2016), the metalicity abundance is $[Fe/H]=-0.08\pm0.07$, leading to a value of metalicity $Z=0.017$ which is close to the Sun metalicity $Z=0:019$.

The positions of the components in the isochrone at
3.612 Gyr indicate that all three components are main
sequence stars. However, the ﬁrst component is transitioning from the main sequence stage to the subgiant stage.
Fragmentation seems to be the most probable mechanism
for the system's formation.

In accordance with Al-Wardat’s method, which is unaffected by variations in parallax measurements (see Table 7),
the most accurate estimate of the system’s mass sum is
$2.39\pm0.18 M_{\odot}$ Al-Wardat et al. (2021b); Hussein et al.
(2023). By combining modiﬁed orbital parameters for the
system with calculated mass sum using Al-Wardat’s
method, we can calculate a new dynamical parallax for
the system using the following equations:
\begin{equation}
3\log{(\pi_{Dyn})}=3\log{h}-log{(\Sigma M)}
\end{equation}
where $\log{(h)}=\log{(a)}-\frac{3}{2}\log{(P)}$. This yields a new dynamical parallax of $(\pi_{Dyn}=23.75\pm0.25 mas)$, close to Hipparcos value $(22.74\pm0.81 mas)$. There is an unusually large
error in Gaia DR2, but DR3 does not resolve the components and does not include a parallax update. We will have
to wait until DR4 is released. Thus, one of the unique
aspects of our analysis is our dynamical parallax, which
is at present the most precise value available.

\begin{figure}
\centering
\includegraphics[width=120mm,height=30mm]{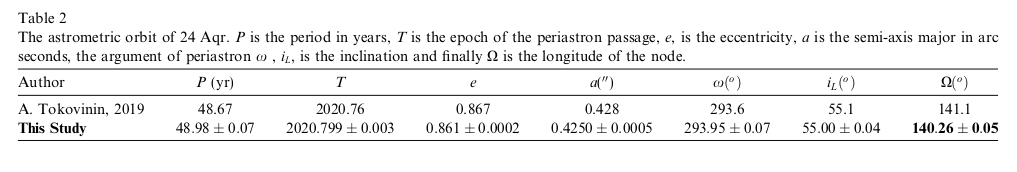}
\end{figure}

\begin{figure}
\centering
\includegraphics[width=120mm,height=40mm]{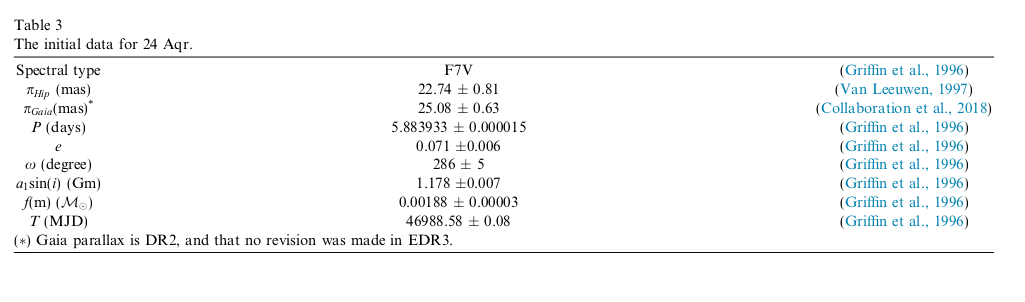}
\end{figure}

\begin{figure}
\centering
\includegraphics[width=160mm,height=90mm]{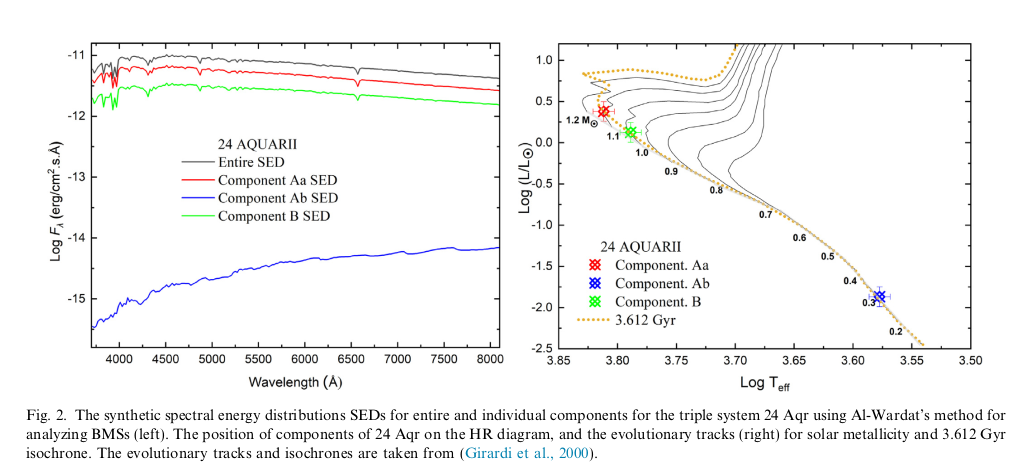}
\end{figure}

\subsection{The Edwards process}

\begin{figure}
\centering
\includegraphics[width=150mm,height=80mm]{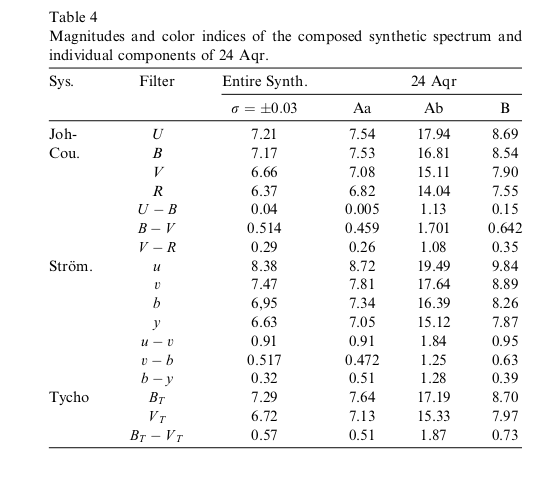}
\end{figure}

\begin{figure}
\centering
\includegraphics[width=140mm,height=70mm]{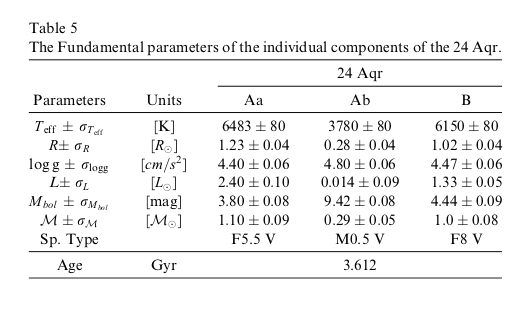}
\end{figure}

\begin{figure}
\centering
\includegraphics[width=140mm,height=40mm]{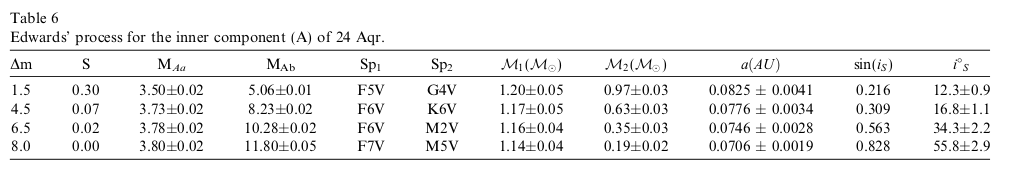}
\end{figure}

\begin{figure}
\centering
\includegraphics[width=140mm,height=40mm]{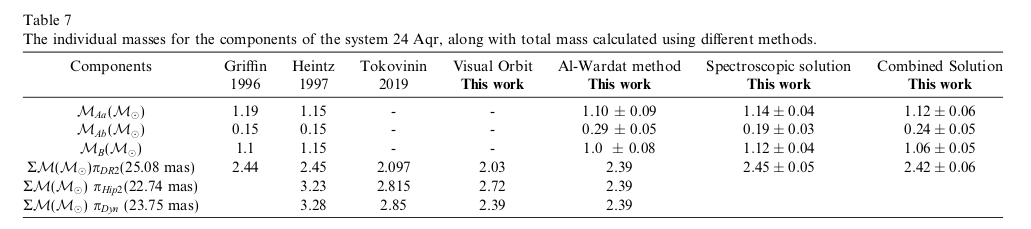}
\end{figure}

Christy and Walker (1969) established individual MK classiﬁcations for both components using MK spectral
class decompositions. As per Terry W. Edwards’s procedure, a standard luminosity calibration for the MK system is employed with a minor modiﬁcation A composite type can be formed by linearly interpolating the MK spectral types of the individual binary components, weighted by
their luminosities (Blaauw and Strand, 1963; Christy and Walker, 1969; Schmidt-Kaler, 1965; Edwards, 1976). This method is of interest for studying the stellar properties of binary stars and their exoplanets. An extremely valuable feature is its simplicity and dependence on apparent size and parallax (Abushattal, 2017; Abushattal et al., 2020; Al-Tawalbeh et al., 2021; Anguita-Aguero et al., 2022; Videla et al., 2022; Chauvin et al., 2022). As a result, the
masses and spectral types of stellar stars are close to those calculated by other astrophysical methods. In this method, we can predict the physical properties of the orbital elements and the orbital diﬀerences. In addition, we can predict the individual mass of spectral binaries by
simultaneously ﬁtting the optical orbit and radial velocity curves. Since the SB1 systems only display the spectral lines of the primary object, the mass of primary object can also
be estimated experimentally (Abushattal et al., 2020; Anguita-Aguero et al., 2022). In the presence of partial and in–homogeneous observations, it provides accurate and efficient estimation of the joint posterior distribution of orbital parameters. The Bayesian inference methodology for parameter estimation is computed by integrating this method with the Markov chain Monte Carlo algorithm (Abushattal et al., 2020). This method was used to calculate
the 3D orbits of SB1, SB2, and conﬁrming data extracted by other methods regarding stellar masses and spectral types. This provided additional external information allowing us to mitigate the problem of not determining the mass ratio as well (Anguita-Aguero et al., 2022). On the other hand, it helps to understand the structures for testing predictions of planet formation and evolution theories in binary systems suc as the structure and stability of the global system and explores HD196885, an extreme planetary system(Chauvin et al., 2022; Algnamat et al., 2022; Alameryeen et al., 2022). Thus, we apply the same method but for spectroscopic binaries then compare the results with any available information about physical or geometrical parameters for the visual binaries such as the total masses of the systems, the inclination, and the angular separation (q) in arc-seconds between the components (see Abushattal, 2017; Abushattal et al., 2020). This method permits the determination of the physical and geometrical
properties using the available data of visual and spectroscopic binaries. It is applicable for the low-mass star with a negligible contribution in the composite spectrum.
Fig. 4 is a graphical representation of the main idea this work. Starting from the visual and spectroscopic observations, it shows all steps required to calculate the physical parameters and orbital elements of the individual components and the most probable 3D orbit. The application of this method permits us to obtain the spectrum of each component from the composite spectrum of the system. We use the following equations (Edwards, 1976; Abushattal, 2017; Abushattal et al., 2020):
\begin{equation}
\ M(S_{1}) + x\;M(S_{2}) = (1+x)\;M(S_{1+2})         
\end{equation}
where $M(S_1), M(S_2), M(S_{1+2})$ are the absolute magnitudes. (S$_{1,2}$) is the absolute magnitude and the visual magnitude difference by :
\begin{equation}
\  \triangle m = M(S_{2}) - M(S_{1}) = M_{2} - M_{1}=-2.5\; log(x)          
\end{equation}

Then Edwards equations can be expressed in the form:
 \begin{equation}
\ M_{1} + x M_{2} = (1+x) M 
\end{equation}
 then we deduce
\begin{equation}
\ M_{1} = M-\frac{x\triangle m}{1+x}. 
\end{equation}

This expression relates ( M$_{1}$) the absolute magnitude of the first component with (M) the total absolute magnitude by means of a step, S.
In this work the global absolute magnitude (M) based on the apparent magnitude and the parallax of the system using well-known relation:
\begin{equation}
\ M = m\;+\;5\;+\;5\;log\;(\pi).
\end{equation}
Therefore, we can represent Edwards step (S) as a function of $\Delta$m for Single-lined (SB1) and double-lined (SB2) spectroscopic binaries:
\begin{equation}
\ S = \frac{x.\triangle m}{1+x}  .  
\end{equation}

Then the composite spectrum is evaluated and so, for
each value of Dm, we can calculate the value of S in Table 6
and then, using the following relation we can calculate the
absolute magnitudes of the both components of the system:
\begin{equation}
\ M_{1}= M - S,  \\
\ M_{2}= M_{1} + \Delta m.
\end{equation} 

The spectral types, the masses and the errors for each of
them can be determined using the calibrations in
(Abushattal et al., 2020 and Pecaut and Mamajek, 2013).
Then we use the value of the mass function from the spectroscopic orbital solution and substitute the previous masses in the following relation in order to determine the
inclination ($i_S$) of the system (see Table 6):
\begin{equation}
i_S=  \ sin \; ^{-1}\left ( \frac{\sqrt[3]{f(\mathcal{M})(\mathcal{M}_{1} + \mathcal{M}_{2})^{2}}}{\mathcal{M}_{2}} \right )  
\end{equation}

Therefore, using data from Table 3 we analyzed the
spectroscopic component (A) in this triple system in order
to describe the physical parameters for Aa and Ab and
compare them with those previously published. So,
depending on the value of the Gaia parallax
(Collaboration et al., 2018), the Edwards process in this
work, and the radial velocity information the most probable value of the masses of 24 Aqr are as follows: the mass of the ﬁrst component (A) is $M_A=1.36\pm05 M_{\odot}$ with
$M_{Aa}=1.14\pm0.04 M_{\odot}$ and $M_{Ab}=0.19\pm0.02 M_{\odot}$ , while the
mass of the second component (B) is $M_B =1.12\pm0.03 M_{\odot}$ . Hence, the total mass of the system is $M_t=2.45\pm0.05M_{\odot}$ (see Table 7).

We employed both the Al-Wardat and Edwards methods, which are observation-based and thus yielded varying results. Consequently, we opted for the average value listed at the table’s end. This average value provides a central measure for the masses, offering simplicity and
comparability.

\section{Orbital alignment}

The system is a hierarchical triple system with well-separated components since the orbital period of the visual
binary is more than 3000 times larger than the one of the
spectroscopic pair. A triple system with such a large period
ratio is dynamically very stable (e.g. Georgakarakos, 2013).
The mutual inclination between the orbital planes of
motion in multistellar systems is an important parameter
in order to understand the system’s formation and evolution (e.g. Sterzik and Tokovinin, 2002; Tokovinin, 1997; Tokovinin, 2014b).

The mutual inclination I between two orbits can be found by (e.g. Fekel, 1981; Docobo, 1977)
\begin{equation}
cos(I)\;\;=\;\; cos (i_{L})\;\; cos (i_{S}) \;\;+\;\; sin(i_{L})\;\;sin (i_{S})\;\; cos \; (\Omega_{L}\;-\;\Omega_{S}),   
\end{equation}
where $i_{L}$ and $i_{S}$ are the orbital inclinations for the outer ($L$) and inner ($S$) orbit respectively, and $\Omega_{L}$ and $\Omega_{S}$ are the corresponding ascending node longitudes.
The spectroscopic binary has $i_{S}=55.8^{o}$, while we have found for the visual binary that $i_{L}=55.0^{o}$ and $\Omega_{L}$ = 140.26$^{o}$. Since the longitude of the ascending node of the inner orbit $\Omega_{S}$ is unknown, the mutual inclination between the two orbits cannot be determined; but it falls within the range  
\begin{equation}
55.8^{\circ}-55.0^{\circ} \leq I \leq 55.8^{\circ}+55.0^{\circ}\Rightarrow  0.8^{\circ} \leq I \leq 110.8^{\circ},
\end{equation}
or
\begin{equation}
124.2^{\circ}-55.0^{\circ} \leq I \leq 124.2^{\circ}+55.0^{\circ}\Rightarrow  69.2^{\circ} \leq I \leq 179.2^{\circ}
\end{equation}
if we consider a retrograde orbit for the spectroscopic binary. The orbital conﬁguration along with the masses of the stars, however, may provide us with some clues so that we can further constrain the mutual inclimation value. Our system constists of a close inner binary with a period of about
5.8 days and a mass ratio m2 =m1 around 0.21 and a third star
at 17.97 au on an elliptic orbit of 0.861 eccentricity. Generally, the formation of close binaries is a complex problem with many pending questions (Kratter, 2011), involving different formation channels (Bate et al., 2002). The formation history of a triple system can provide information regarding
the orientation of the orbital planes of the system. According
to Bate (2009), the formation of a triple system from disc
fragmentation around an initially single object or a ﬂatten
core, favours aligned orbital planes. In addition, the accretion of mass in a triple system that forms with initially non-coplanar orbital planes may result in a decrease of the mutual inclination of that system into closer alignment.
Finally, triple systems formed from single object capture
by a binary may have their orbital planes very misaligned.
A passing object can have a similar eﬀect on the alignment of the orbital planes of a triple system.

One possible mechanism of producing short period
binaries in hierarchical triple systems is Kozai cycles with
tidal friction (KCTF) (e.g. Fabrycky and Tremaine, 2007). As noticed by Kozai (1962) and Lidov (1962), a
highly inclined ($39.2 ^{\circ} < I < 140.8 ^{\circ}$) tertiary companion
with respect to the orbital plane of a binary orbit would
push the eccentricity of the binary to high values, even to
1 for a perpendicular conﬁguration. As the high eccentricity values lead to close pericentre passages for the two
members of the inner binary, tidal friction becomes important and works towards the shrinking and circularisation of the binary orbit. Other effects, such as for example the relativistic precession of the pericentre of the binary orbit, are also expected to contribute to the orbital evolution of the
system. In their numerical experiments, Fabrycky and
Tremaine (2007) found that KCTF produced triple systems
with very large period ratios, circular inner orbits, and a
mutual inclination that was often near the critical one.
Moe and Kratter (2018), however, suggested that only a
fraction of close binaries can be explained through the
KCTF mechanism. They also stated that $60\%$ of close binaries with orbital periods less than 10 days derive from the
dynamical unfolding of initially unstable triples that fragment in the disk coupled with significant energy dissipation
within the disk. The close binaries formed that way have
almost coplanar tertiary components in conﬁgurations
where the outer semi-major axis is between 0.5 and 50
au. This formation channel operates exclusively during the
pre-MS phase while there is still dissipative gas in the primordial disk. Hence, interactions of coplanar triples
embedded in disks, not secular evolution of misaligned triples nor disk migration of solitary binaries, may explain the
majority of very close binaries. Similar formation scenarios
are discussed in Tokovinin (2020) and Tokovinin and Moe
(2020).

Moreover, Bate (2012); Bate (2014), after performing
hydrodynamical simulations of star cluster formation, concluded that both observed and simulated triple systems
have a tendency towards orbital coplanarity. This conclusion seems to be supported by Tokovinin (2017), where
he found a rather strong tendency of orbit alignment in triple stars where the outer separation was less than 50 au. He
also found that the orbit alignment was stronger in triple
stars with low-mass primaries. It is also, however, pointed
out in that work that gas accretion by the stars can create
misaligned triples in systems with randomly aligned angular momentum at the epoch of star formation. Along with
changes in the outer orbit orientation, it can also cause the
rapid inward migration of the outer body. This may
increase the intensity of the gravitational perturbations
among the bodies of the system, which in certain cases
may even lead to its destabilization. Such kind of dynamical interactions in multi-body systems could result in misaligned triples and often with highly eccentric orbits
(Antognini and Thompson, 2016). On the other hand, triple stars produced in N-body decay, with or without accretion, usually have moderate period ratios
$(\approx 10)$ and outer mass ratios $m_3/(m_1+m_2)<0.2$ (Tokovinin, 2008).

Griffin et al. (1996) gave a 0.071 eccentricity for the
spectroscopic binary orbit. Assuming coplanar orbits, from
the work of Georgakarakos (2003), the forced eccentricity
of the spectroscopic binary orbit is
\begin{equation}
e_{forced}=\frac{C}{B-A}\approx 0.0150,
\end{equation}
where 

\begin{eqnarray}
A = \frac{\mathcal{M}_{Aa}\mathcal{M}_{Ab}\mathcal{M}_t^{1/2}}{\mathcal{M}_B\mathcal{M}_A^{3/2}}\left(\frac{a_S}{a_L}\right)^{1/2}\frac{1}{(1-e_L^2)^2}\nonumber\\
B = \frac{1}{(1-e_L^2)^{3/2}}+\frac{25}{8}\frac{\mathcal{M}_{B}}{\mathcal{M}_t^{1/2}\mathcal{M}_A^{1/2}}\left(\frac{a_S}{a_L}\right)^{3/2}\frac{3+2e_L^2}{(1-e_L^2)^3}\nonumber\\
C = \frac{5}{4}\frac{\mathcal{M}_{Aa}-\mathcal{M}_{Ab}}{\mathcal{M}_{A}}\frac{a_S}{a_L}\frac{e_L}{(1-e_L^2)^{5/2}},\nonumber
\end{eqnarray}
with $\mathcal{M}_{Aa}=1.12 \mathcal{M_{\odot}}$, $\mathcal{M}_{Ab}=0.24 \mathcal{M_{\odot}}$, $\mathcal{M}_{B}=1.06 \mathcal{M_{\odot}}$,
$a_S=0.0706 au$, $a_L=17.97 au$ and $e_L=0.861$.

If the spectroscopic orbit is initially circular, its maximum orbital eccentricity will reach the value of 0.030, i.e. twice the value of the forced eccentricity but still well below our current 0.071. Of course, that maximum value can be higher if we consider that the initial eccentricity of the spectroscopic binary is not zero. Considering a mutually
inclined orbit could easily provide the value we observe as long as 
$i>39.2^{\circ}$. We have to point out here that since
the semi-major axis of the inner binary is 0.0706 au, other
effects, such as tidal friction, deformation of the shape of
the stars due to rotation, and general relativity may contribute to the evolution of the inner orbit by suppressing
Kozai cycles (Fabrycky and Tremaine, 2007; Taani et al.,
2020). But even at low mutual inclinations or coplanar
orbits the result will be the same, i.e. the eccentricity oscillations will be supressed resulting in lower amplitudes.  Fig. 3 demonstates that eﬀect for diﬀerent values of the
mutual inclimation $i$. In those plots, we have only taken
into consideration the general relativistic eﬀect which,
based on some crude estimates using the equations given
in Fabrycky and Tremaine (2007), seems to be the dominant one here. A quick calculation yields an order of magnitude of $10^{-4}$ for the rate of the relativistic pericentre precession, while the precession rate due to tidal distortion is an order of magnitude less, i.e. $10^{-5}$ . Finally, the precession rate due to the rotational distortion of the stars is several orders of magnitude smaller than the other two effects.
Only in the case of a highly inclined outer orbit we can
achieve the value of the eccentricity of the spectroscopic
orbit as given in Griffin et al. (1996). More speciﬁcally, taking into consideration general relativistic eﬀects, the mutual
inclination of the system needs to be above $60^{\circ}$ and less
than $120^{\circ}$ if we want to achieve a spectroscopic binary
eccentricity of about 0.071.

To summarise, the mutual inclination of the system
seems to be in the range $[0.8^{\circ},110.8^{\circ}]$ or $[69.2^{\circ},179.2^{\circ}]$ i.e. the system could be nearly coplanar, highly inclined or having the distant star to move on a retrograde orbit with respect to the spectroscopic binary. There are certain characteristics regarding the orbits and masses of the system that may point towards the alignment or not of the two orbital planes. On the other hand, it appears that the observed value of 0.071 for the eccentricity of the spectroscopic binary is easier to be achieved by Kozai-Lidov oscillations, even though they are suppressed by other physical processes.

\begin{figure}
\centering
\includegraphics[width=150mm,height=180mm]{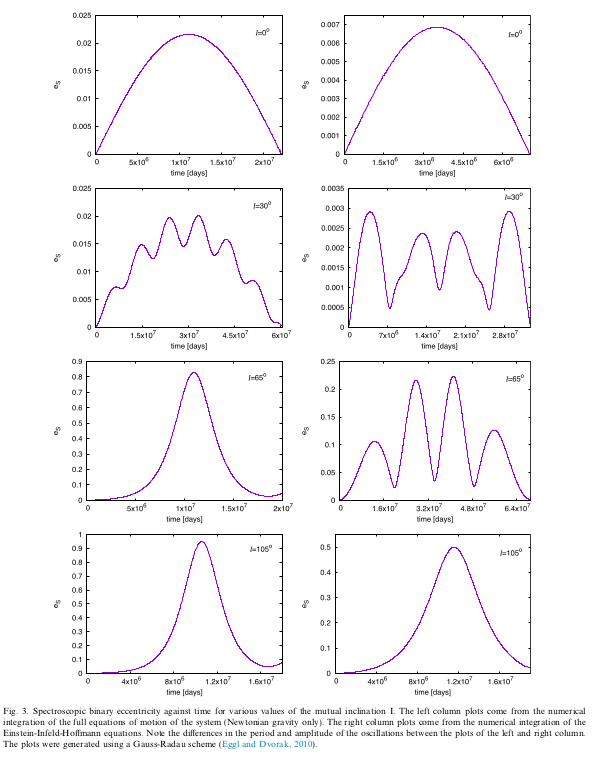}
\end{figure}

\begin{figure}
\centering
\includegraphics[width=140mm,height=160mm]{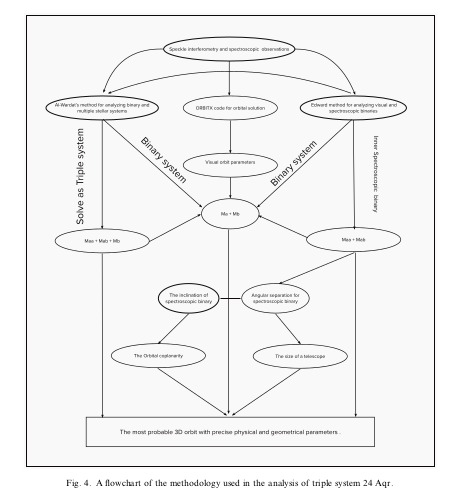}
\end{figure}

\section{Is it possible to resolve this system visually?}

In 2021, the wide system passed through the periastron
passage, which helped us determine the dips, rotational
velocities, and mass ratios of the system’s components. A
valid question to answer is whether it is possible to resolve
the entire 24 Aqr system visually. In other words, what is
the size of the telescope we need to resolve the inner system? Previously, we determined the physical and orbital
elements of the entire system. We have all the required
parameters to calculate the ephemerides of the inner spectroscopic binary by the algorithm for visual binaries. We
only need to describe the position of the inner binary components in the apparent plane where the polar coordinates
$\theta$ and $\rho$ are deﬁned. We will measure the position angle $\theta$
from the line of the nodes. Then we can describe the relative orbit based on two-body formulae and we determine
the coordinates $r$ and $f$ for an epoch $t$. We calculate the separation between the components of the inner binary by estimating the mean anomaly, $l$, then calculate the eccentric anomaly, $E$, using the following equation from
(Abushattal, 2017; Abushattal et al., 2020):
\begin{equation}
\label{3law123}
\ r''= a'' \:(1- e\:cos\:(E)\:)
\end{equation}
\noindent where the double prime denotes arcsecs, and the true anomaly,
\begin{equation}
\label{3law124}
\ tan\:(\frac{\textit{f}}{2})=\sqrt{\frac{1+e}{1-e}}\:tan\,(\frac{E}{2}) .
\end{equation}	
Then

\begin{equation}
\ tan\:(\theta-\Omega)=\:tan\:(\omega+\textit{f}\:)\:cos i
\end{equation}
and
\begin{equation}
\ \rho''= r''\: \frac{cos\: (\omega+\textit{f}\:)}{cos\:(\theta-\Omega)}. 
\end{equation}

\noindent $\omega$ is the longitude of the periastron, and we assume that $\Omega$ = 0$^{\circ}$. Then we can calculate $\theta$ and $\rho^{''}$. More details about the above calculations can be found in Abushattal et al. (2020).  The apparent orbit can be drawn sufficiently using  $0^{\circ}$ $\leq$ $\theta$ $\leq$ $360^{\circ}$ in (18).  $f$ can be calculated using same expression (18). Then we can describe the minimum and maximum separation $\rho$ between the components of the inner orbit.

The ability of the telescope to resolve the binary stars
visually depends on many factors such as relative brightness,lens and mirror coatings, the condition of sky, and light pollution even the observer experience. ‘‘The Double Double” star, also known as Epsilon Lyrae in the constellation of Lyra, is an example for the relation between the
power separation between the components and the increase
of the size of the telescopes; Dawes limit describes that with
the following formula (Mullaney, 2006):
\begin{equation}
\ D = \frac{11.7}{\rho''}, 
\end{equation}
where the angular separation $\rho$ is in arc seconds and  the diameter D in centimeters. Hence, we use the maximum value for $\rho$ derived by the  equations above and which equals to $0.0018^{''}$. In this case, we need a telescope with a diameter larger than 65 meters to resolve the inner spectroscopic binary in 24 AQUARII triple system. We suggest that the system should be observe with a large array of telescopes such as CHARA (Center for High Angular Resolution Astronomy) array and Navy Precision Optical Interferometer (NPOI)  which have resolving capability larger than 300 meters as single telescope. Also, we can put this system as a test for future high resolution monitoring tools. 

\section{Coclusions}
We present a spectro-interferometric analysis, atmo-
spheric modeling and synthetic photometry for the hierarchical triple System 24 Aqr. We used new relative position measurements to modify the orbit of the outer
system and to ﬁnd new orbital elements and dynamical
parallax. Three diﬀerent methods were used in the analysis, these are: Tokovinin’s dynamical method (ORBITX) for calculating the orbit of the outer binary. Edwards’ method for solving the inner single-line spectroscopic binary. Al-Wardat’s method for analyzing binary and multiple stellar systems for building synthetic SEDs for each of the the three components, and for estimating the complete set of physical and geometrical
parameters of the system, including individual masses,
synthetic photometry and a new parallax for the system.
The main conclusions of this work can be summarized as
follows:
\begin{itemize}
\item A modified orbital solution, and new orbital elements were calculated for the outer binary using four new speckle-interferometric measurements.
\item We estimated the individual masses of the system using a combination of different methods. Edwards' method gives $\mathcal{M}_{Aa}$= 1.14$\pm$0.04$\mathcal{M}_{\odot}$ and $\mathcal{M}_{Ab}$ = 0.19$\pm$0.03 $\mathcal{M}_{\odot}$ and $\mathcal{M}_{B}$ = 1.12$\pm$0.04 $\mathcal{M}_{\odot}$.
 Al-Wardat's method method gives $\mathcal{M}_{Aa}$= $1.10\pm0.09$ $\mathcal{M}_{\odot}$ and $\mathcal{M}_{Ab}$ = $0.29\pm0.05$ $\mathcal{M}_{\odot}$ and $\mathcal{M}_{B}$ = $ 1.0\pm0.08$ $\mathcal{M}_{\odot}$.

\item   We introduced  accurate atmospheric and fundamental parameters for  the individual components. These parameters include  the individual masses, the effective temperatures, gravity, stellar radii, luminosity, magnitudes, color indices, spectral types and ages. 
\item According to Al-Wardat's method, which barely affected by any change in the parallax measurements, as it is clear in Table 7, and as it was concluded by  Al-Wardat et al. (2021b), the best estimation for the mass sum of the system is ($2.74\pm0.18 \mathcal{M}_{\odot}$).
This leads us to adopt a new dynamical parallax  for the system as  $\pi_{Dyn} (\pi=22.63\pm 0.25$ mas), which is very close to that given by Hipparcos 2 as $22.74\pm0.81$ mas. 
DR2 has an atypically large error, DR3 has no parallax update and neither resolves the components, and we will have to wait to see about DR4. Thus, our dynamical parallax is in fact the most precise value that exists at present, it is one of the unique aspects of our analysis.

\item We discussed the orbital alignment of the two orbits in this hierarchical stellar triple. 

\item 
Finally, we gave the required specifications of the telescope to resolve the inner spectroscopic system interferometrically.    
      
\end{itemize}

\section*{Data availability}
We will publish the new measurements used in this calculation in this article. The stellar data and other measurements were obtained from the data bases and catalogs mentioned in the acknowledgments. In response to reasonable requests, the corresponding author will provide the data underlying this article.

\section*{Declaration of Competing Interest}
The authors declare that they have no known competing
ﬁnancial interests or personal relationships that could have
appeared to inﬂuence the work reported in this paper.

\section*{Acknowledgments}
Thank Gerard van Belle, Catherine Clarck, and Zachary
Hartman at Lowell Observatory for helping us obtain the
QWSSI observations. We are grateful to US Naval Observatory for both the Fourth Catalog of Interferometric
Observations of Binary Stars and the Washington Double
Star Catalog. Also, to Centre de donnees astronomiques de
Strasbourg (CDS), Strasbourg,France for the available
SIMBAD database. This work has made use of data from
the European Space Agency (ESA) mission Gaia(https://
www.cosmos.esa.int/gaia), processed by the Gaia Data
Processing and Analysis Consortium (DPAC, https://
www.cosmos.esa.int/web/gaia/dpac/consortium). It also
has made use of SIMBAD database, the Fourth Catalog
of Interferometric Measurements of Binary Stars, IPAC
datasystems, and codes of Al-Wardat’s method for analyzing binary and multiple stellar systems.

\section*{References}
Abt, H., 1985. Visual multiples. viii-1000 mk types. Astrophys. J. Suppl.
Ser. 59, 95–112.\\
Abu-Alrob, E.M., Hussein, A.M., Al-Wardat, M.A., 2023. Atmospheric
and fundamental parameters of the individual components of multiple
stellar systems. Astronom. J. 165 (6), 221. https://doi.org/10.3847/
1538-3881/acc9ab.\\
Abushattal, A., Al-Wardat, M., Taani, A., Khassawneh, A., Al-Naimiy,
H., 2019. Extrasolar planets in binary systems (statistical analysis).
Journal of Physics: Conference Series, vol. 1258. IOP Publishing, p.
012018.\\
Abushattal, A., Alrawashdeh, A., Kraishan, A., 2022a. Astroinformatics:
The importance of mining astronomical data in binary stars cata-
logues. Commun. BAO 69 (2), 251–255.\\
Abushattal, A., Kraishan, A., Alshamaseen, O., 2022b. The exoplanets
catalogues and archives: An astrostatistical analysis. Commun. BAO
69 (2), 235–241.\\
Abushattal, A.A., 2017. The modeling of the physical and dynamical
properties of spectroscopic binaries with an orbit: doctoral disserta-
tion. Universidade de Santiago de Compostela, Ph.D. thesis.\\
Abushattal, A.A., Docobo, J.A., Campo, P.P., 2020. The most probable
3D Orbit for spectroscopic binaries. Astronom. J. 159 (1), 28. https://
doi.org/10.3847/1538-3881/ab580a.\\
Aitken, R., Millard, R.E., 1932. The orbit of the binary system b1212= 24
aquarii. Publ. Astron. Soc. Pac. 44 (258), 129–131.\\
Aitken, R.G., 1932. The Orbit of the binary system b 1212=24. Aquarii.
44 (258), 129.\\https://doi.org/10.1086/124215.\\
Al-Tawalbeh, Y.M., Hussein, A.M., Taani, A., Abushattal, A., Yusuf, N.,
Mardini, M., Suleiman, F., Al-Naimiy, H.M., Khasawneh, A.M., Al-
Wardat, M.A., 2021. Precise masses, ages, and orbital parameters of
the binary systems hip 11352, hip 70973, and hip 72479. Astrophys.
Bull. 76, 71–83.\\
Al-Wardat, M., 2012. Physical parameters of the visually close binary
systems Hip70973 and Hip72479. Publ. Astronom. Soc. Aust. 29 (4),
523–528. https://doi.org/10.1071/AS12004, arXiv:1204.4589.\\
Al-Wardat, M., Docobo, J., Abushattal, A., Campo, P., 2017. Physical
and geometrical parameters of cvbs. xii. ﬁn 350 (hip 64838). Astrophys.
Bull. 72(1), 24–34.\\
Al-Wardat, M.A., 2002. Spectral energy distributions and model atmo-
sphere parameters of the quadruple system ADS11061. Bull. Special
Astrophys. Obser. 53, 51–57.\\
Al-Wardat, M.A., 2003a. Model atmosphere parameters of the binary
system 41 Dra. Bull. Special Astrophys. Obser. 56, 41–45.\\
Al-Wardat, M.A., 2003b. Spectral energy distributions and model
atmosphere parameters of the binary systems COU1289 and
COU1291. Bull. Special Astrophys. Obser. 56, 36–40.\\
Al-Wardat, M.A., 2007. Model atmosphere parameters of the binary
systems COU1289 and COU1291. Astron. Nachr. 328 (1), 63–67.
https://doi.org/10.1002/asna.200610676.\\
Al-Wardat, M.A., 2008. Synthetic photometry of speckle interferometric
binaries. Astrophys. Bull. 63 (4), 361–365. https://doi.org/10.1134/
S1990341308040044.\\
Al-Wardat, M.A., Abu-Alrob, E., Hussein, A.M., Mardini, M.K., Taani,
A.A., Widyan, H.S., Yousef, Z.T., Al-Naimiy, H.M., Yusuf, N.A.,
2021a. Physical and geometrical parameters of CVBS XIV: the two
nearby systems HIP 19206 and HIP 84425. Res. Astron. Astrophys. 21
(7),161. https://doi.org/10.1088/1674-4527/21/7/161, arXiv:2111.11675.\\
Al-Wardat, M.A., Hussein, A.M., Al-Naimiy, H.M., Barstow, M.A.,
2021b. Comparison of Gaia and Hipparcos parallaxes of close visual
binary stars and the impact on determinations of their masses. Publ.
Astronom. Soc. Aust. 38, e002. https://doi.org/10.1017/pasa.2020.50,
arXiv:2111.05325.\\
Al-Wardat, M.A., Widyan, H., 2009. Parameters of the visually close
binary system Hip11253 (HD14874). Astrophys. Bull. 64 (4), 365–371.
https://doi.org/10.1134/S1990341309040063.\\
Al-Wardat, M.A., Widyan, H.S., Al-thyabat, A., 2014. Complex analysis
of the stellar binary HD25811: a subgiant system. Publ. Astronom.
Soc.Aust. 31, e005. https://doi.org/10.1017/pasa.2013.42, arXiv:1311.5721.\\
Alameryeen, H., Abushattal, A., Kraishan, A., 2022. The physical
parameters, stability, and habitability of some double-lined spectro-
scopic binaries. Commun. BAO 69 (2), 242–250.\\
Algnamat, B., Abushattal, A., Kraishan, A., Alnaimat, M., 2022. The
precise individual masses and theoretical stability and habitability of
some single-lined spectroscopic binaries. Commun. BAO 69 (2), 223–
230.\\
Anguita-Aguero, J., Mendez, R.A., Claverıia, R.M., Costa, E., 2022.
Orbital elements and individual component masses from joint
spectroscopic and astrometric data of double-line spectroscopic
binaries. Astron. J. 163 (3), 118.\\
Antognini, J.M.O., Thompson, T.A., 2016. Dynamical formation and
scattering of hierarchical triples: cross-sections. Kozai-Lidov Oscilla-
tions Collisions. 456 (4), 4219–4246. https://doi.org/10.1093/mnras/
stv2938, arXiv:1507.03593.\\
Balega, Y.Y., Ryadchenko, V., 1984. Digital speckle interferometry of
binary stars. Soviet Astron. Lett. 10, 95–98.\\
Bate, M.R., 2009. Stellar, brown dwarf and multiple star properties from
hydrodynamical simulations of star cluster formation., 392(2), 590–
616. https://doi.org/10.1111/j.1365-2966.2008.14106.x. arXiv:0811.0163.\\
Bate, M.R., 2012. Stellar, brown dwarf and multiple star properties from a
radiation hydrodynamical simulation of star cluster formation. 419(4),
3115–3146. https://doi.org/10.1111/j.1365-2966.2011.19955.x. arXiv:1110.1092.\\
Bate, M.R., 2014. The statistical properties of stars and their dependence
on metallicity: the eﬀects of opacity. 442(1), 285–313. https://doi.org/
10.1093/mnras/stu795. arXiv:1405.5583.\\
Bate, M.R., Bonnell, I.A., Bromm, V., 2002. The formation of close
binary systems by dynamical interactions and orbital decay. 336(3),
705–713. https://doi.org/10.1046/j.1365-8711.2002.05775.x. arXiv:astro-ph/0212403.\\
Blaauw, A., Strand, K.A., 1963. Basic astronomical data. Aa. Strand, Ed,
p. 383.\\
Bonneau, D., Blazit, A., Foy, R., Labeyrie, A., 1980. Speckle interfero-
metric measurements of binary stars. Astron. Astrophys. Suppl. Ser.
42, 185–188.\\
Cannon, A., Pickering, E., 1924. Henry draper (hd) catalog and hd
extension. hdhc.\\
Carter, J.A., Fabrycky, D.C., Ragozzine, D., Holman, M.J., Quinn, S.N.,
Latham, D.W., Buchhave, L.A., Van Cleve, J., Cochran, W.D., Cote,
M.T., et al., 2011. Koi-126: A triply eclipsing hierarchical triple with
two low-mass stars. Science 331 (6017), 562–565.\\
Chauvin, G., Videla, M., Beust, H., Mendez, R., Correia, A., Lacour, S.,
Tokovinin, A., Hagelberg, J., Bouchy, F., Boisse, I. et al., 2022.
Chasing extreme planetary architectures: I-hd196885ab, a super-jupiter
dancing with two stars? arXiv preprint arXiv:2211.00994.\\
Christy, J.W., Walker Jr, R., 1969. Mk classiﬁcation of 142 visual binaries.
Publications of the Astronomical Society of the Paciﬁc, pp. 643–649.\\
Clark, C.A., Van Belle, G.T., Horch, E.P., Trilling, D.E., Hartman, Z.D.,
Collins, M., Von Braun, K., Gehring, J., 2020. The optomechanical
design of the quad-camera wavefront-sensing six-channel speckle
interferometer (qwssi). In: Optical and Infrared Interferometry and
Imaging VII, vol. 11446, SPIE, pp. 540–548.\\
Collaboration, G. et al., 2018. Vizier online data catalog: Gaia dr2 (gaia
collaboration, 2018). yCat, pp. I–345.\\
Danjon, A., 1942. Orbites de 24 aquarii, a 88 et beta 612. J. des Obser. 25,
18.\\
Derekas, A., Kiss, L.L., Borkovits, T., Huber, D., Lehmann, H.,
Southworth, J., Bedding, T.R., Balam, D., Hartmann, M., Hrudkova,
M., et al., 2011. Hd 181068: A red giant in a triply eclipsing compact
hierarchical triple system. Science 332 (6026), 216–218.\\
Docobo, J., Balega, Y., Campo, P., Abushattal, A., 2018a. Double stars
inf.\\
Docobo, J., Campo, P., Abushattal, A., 2018b. Iau commiss. Double Stars
169, 1.\\
Docobo, J.A., Andrade, M., 2006. A methodology for the description of
multiple stellar systems with spectroscopic subcomponents. Astrophys.
J. 652 (1), 681.\\
Docobo, J.A., Griﬃn, R.F., Campo, P.P., Abushattal, A.A., 2017. Precise
orbital elements, masses and parallax of the spectroscopic–interferometric binary hd 26441. Mon. Not. R. Astron. Soc. 469 (1), 1096–1100.\\
Edwards, T., 1976. Mk classiﬁcation for visual binary components.
Astron. J. 81, 245–249.\\
Eggen, O.J., 1963. The empirical mass-luminosity relation. Astrophys. J.
Suppl. Ser. 8, 125.\\
Eggl, S., Dvorak, R., 2010. An introduction to common numerical
integration codes used in dynamical astronomy. In: Souchay, J.,
Dvorak, R.(Eds.), Lecture Notes in Physics, vol. 790, Berlin Springer
Verlag, https://doi.org/10.1007/978-3-642-04458-8${\_}$9, pp. 431–480.\\
Fabrycky, D., Tremaine, S., 2007. Shrinking binary and planetary orbits
by kozai cycles with tidal friction. Astrophys. J. 669 (2), 1298.\\
Fekel, F.C., Scarfe, C., Barlow, D., Duquennoy, A., McAlister, H.A.,
Hartkopf, W., Mason, B., Tokovinin, A., 1997. New and improved
parameters of hd 202908= ads 14839: a spectroscopic-visual triple
system. Astron. J. 113, 1095.\\
Fekel Jr, F., 1981. The properties of close multiple stars. Astrophys. J.
246, 879–898.\\
Finsen, W., 1929a. The orbit of 24 aquarii (beta 1212, bu. gc 11125). 23h
34m 4,-0 30’(1900). Circ. Union Obser. Johannesburg 81, 112–114.\\
Finsen, W.S., 1929b. The Orbit of 24 Aquarii (b 1212, Bu. G.C. 11125).
23h 34m4, -0 30’ (1900). Circ. Union Obser. Johannesburg 81, 112–
114.\\
Gaspar, A., Rieke, G.H., Ballering, N., 2016. The correlation between
metallicity and debris disk mass. Astrophys. J. 826 (2), 171. https://doi.
org/10.3847/0004-637X/826/2/171, arXiv:1604.07403.\\
Georgakarakos, N., 2002. Eccentricity generation in hierarchical triple
systems with coplanar and initially circular orbits. Mon. Not. R.
Astron. Soc. 337 (2), 559–566.\\
Georgakarakos, N., 2003. Eccentricity evolution in hierarchical triple
systems with eccentric outer binaries. 345(1), 340–348. https://doi.org/
10.1046/j.1365-8711.2003.06942.x. arXiv:1408.5890.\\
Georgakarakos, N., 2005. Erratum: Eccentricity evolution in hierarchical
triple systems with eccentric outer binaries. Mon. Not. R. Astron. Soc.
362 (2), 748–748.\\
Georgakarakos, N., 2013. The dependence of the stability of hierarchical
triple systems on the orbital inclination. 23, 41–48. https://doi.org/10.
1016/j.newast.2013.02.004. arXiv:1302.5599.\\
Girardi, L., Bressan, A., Bertelli, G., Chiosi, C., 2000. Evolutionary tracks
and isochrones for low -and intermediate- mass stars: From 0.15 to 7
Msun, and from Z=0.0004 to 0.03. Astronom. Astrophys. Suppl. Ser.
141, 371–383. https://doi.org/10.1051/aas:2000126, arXiv:astro-ph/
9910164.\\
Girardi, L., Bressan, A., Bertelli, G., Chiosi, C., 2000. Vizier online data
catalog: Low-mass stars evolutionary tracks and isochrones (girardi+,
2000). VizieR Online Data Catalog, pp. J-A+.\\
Griffin, R., Carquillat, J., Ginestet, N., Udry, S., 1996. Spectroscopic
binary orbits from photoelectric radial velocities. paper 128: 24
aquarii. Observatory 116, 162–175.\\
Harlan, E., 1974. Mk classiﬁcation for f-and g-type stars. iii. Astron. J. 79,
682–686.\\
Hartkopf, W., Mason, B., Wycoﬀ, G., McAlister, H., 2010. Fourth
catalogue of interferometric measurements of binary stars, http://ad.
usno.navy.mil/wds/int4.html(ic4).\\
Heintz, W., 1981. Radial velocities of binary and proper-motion stars.
Astrophys. J. Suppl. Ser. 46, 247.\\
Heintz, W.D., 1997. The triple star 24 aquarii. Observatory 117, 93–93.
Hussein, A.M., Abu-Alrob, E.M., Mardini, M.K., Alslaihat, M.J., Al-Wardat, M.A., 2023. Complete analysis of the subgiant stellar system:
Hip 102029. Adv. Space Res.\\
Hussein, A.M., Al-Wardat, M.A., Abushattal, A., Widyan, H.S., Abu-
Alrob, E.M., Malkov, O., Barstow, M.A., 2022. Atmospheric and
fundamental parameters of eight nearby multiple stars. Astronom. J.
163 (4), 182. https://doi.org/10.3847/1538-3881/ac4fc7.\\
JA Docobo, D., 1977. Aplicacion de la teoria de perturbaciones al estudio
de sistemas estelares triples. Ph.D. thesis Universidad de Zaragoza.\\
Kozai, Y., 1962. Secular perturbations of asteroids with high inclination
and eccentricity. Astronom. J. 67, 591–598. https://doi.org/10.1086/
108790.\\
Kratter, K.M., 2011. The Formation of Close Binaries. In: Schmidtobre-
ick, L., Schreiber, M.R., Tappert, C. (Eds.), Evolution of Compact
Binaries, vol. 447, Astronomical Society of the Paciﬁc Conference
Series, p. 47. arXiv:1109.3740.\\
Kuiper, G., 1926. The orbit of bu. gc 11125= beta1212= 24 aquarii. Bull.
Astron. Inst. Netherlands 3, 147.\\
Kurucz, R., 1994a. Atomic Data for Ca, Sc, Ti, V, and Cr. Atomic Data
for Ca, 20.\\
Kurucz, R., 1994b. Atomic Data for Fe and Ni. Atomic Data for Fe and
Ni. Kurucz CD-ROM No. 22. Cambridge, 22.\\
Kurucz, R., 1994c. Atomic Data for Mn and Co., Atomic Data for Mn
and Co. Kurucz CD-ROM No. 21. Cambridge, 21.\\
Kurucz, R., 1994d. Solar abundance model atmospheres for 0,1,2,4,8 km/
s. Solar abundance model atmospheres for 0, 19.\\
Kurucz, R.L., 1994e. Computation of Opacities for Diatomic Molecules.
In: Jorgensen, U.G. (Ed.), IAU Colloq. 146: Molecules in the Stellar
Environment, vol. 428. p. 282, https://doi.org/10.1007/3-540-57747-5$\_$51.\\
Lidov, M.L., 1962. The evolution of orbits of artiﬁcial satellites of planets
under the action of gravitational perturbations of external bodies. 9
(10), 719–759. https://doi.org/10.1016/0032-0633(62)90129-0.\\
Lippincott, S., 1982. Masses mass ratios and parallaxes from astrometric
studies of seven visual binaries. Astron. J. 87, 1237.\\
Malkov, O.Y., 2007. Mass–luminosity relation of intermediate-mass stars.
Mon. Not. R. Astron. Soc. 382 (3), 1073–1086.\\
Mannino, G., 1946. Orbita provvisoria delle doppie visuali ADS 15176 =
24 Aquarii ed ADS 15267 = H0 166, 18, 133.\\
Mannino, G., Humblet, J., 1955. Observations spectroscopiques de
quelques étoiles of (i). In: Annales d’Astrophysique, vol. 18, p. 237.\\
Masda, S.G., Docobo, J.A., Hussein, A.M., Mardini, M.K., Al-Ameryeen, H.A., Campo, P.P., Khan, A.R., Pathan, J.M., 2019. Physical and dynamical parameters of the triple stellar system: HIP109951. Astrophys. Bull.. 74 (4), 464–474. https://doi.org/10.1134/S1990341319040126, arXiv:1911.09972.\\
Mendez, R., Tokovinin, A., Horch, E., 2018. A speckle survey of southern
hipparcos visual doubles and geneva-copenhagen spectroscopic bina-
ries. RMxAC 50, 56–57.\\
Moe, M., Kratter, K.M., 2018. Dynamical formation of close binaries
during the pre-main-sequence phase. Astrophys. J. 854 (1), 44. https://
doi.org/10.3847/1538-4357/aaa6d2, arXiv:1706.09894.\\
Mullaney, J., 2006. Double and Multiple Stars, and How to Observe Them.
Springer Science and Business Media.\\
Naoz, S., Farr, W.M., Lithwick, Y., Rasio, F.A., Teyssandier, J., 2013.
Secular dynamics in hierarchical three-body systems. Mon. Not. R.
Astron. Soc. 431 (3), 2155–2171.\\
Pecaut, M.J., Mamajek, E.E., 2013. Intrinsic colors, temperatures, and
bolometric corrections of pre-main-sequence stars. Astrophys. J.
Suppl. Ser. 208 (1), 9.\\
Perryman, M., Lindegren, L., Kovalevsky, J., Hoeg, E., Bastian, U.,
Bernacca, P., Crézé, M., Donati, F., Grenon, M., Grewing, M. et al.,
1997. The hipparcos catalogue.\\
Prusti, T., De Bruijne, J., Brown, A.G., Vallenari, A., Babusiaux, C.,
Bailer-Jones, C., Bastian, U., Biermann, M., Evans, D., Eyer, L., et al.,
2016. The gaia mission. Astron. Astrophys. 595, A1.\\
Scardia, M., Prieur, J.-L., Pansecchi, L., Ling, J., Argyle, R., Aristidi, E.,
Zanutta, A., Abe, L., Bendjoya, P., Rivet, J.-P. et al., 2019. Orbital
elements of double stars: Ads 12648, ads 12889. Information Circular-
IAU Commission G1 Double Stars, 199, 3–4.\\
Schmidt-Kaler, T., 1965. Landolt-bornstein, kh hellwege, ed. Berlin:
Springer-Verlag, New Ser., Group, 6, 301.\\
Stassun, K.G., 2012. Astrophysics: A pas de trois birth for wide binary
stars. Nature 492 (7428), 191–192.\\
Sterzik, M.F., Tokovinin, A.A., 2002. Relative orientation of orbits in
triple stars. Astron. Astrophys. 384 (3), 1030–1037.\\
Straizys, V., Kuriliene, G., 1981. Fundamental stellar parameters derived
from the evolutionary tracks. Astrophys. Space Sci. 80 (2), 353–368.\\
Taani, A., Abushattal, A., Khasawneh, A., Almusleh, N., Al-Wardat, M.,
2020. Jordan journal of physics. Jordan J. Phys. 13 (3), 243–251.\\
Taani, A., Abushattal, A., Mardini, M.K., 2019a. The regular dynamics
through the ﬁnite-time lyapunov exponent distributions in 3d hamiltonian systems. Astron. Nachr. 340 (9–10), 847–851.\\
Taani, A., Karino, S., Song, L., Mardini, M., Al-Wardat, M., Abushattal,
A., Khasawneh, A., Al-Naimiy, H., 2019b. On the wind accretion
model of gx 301–2. Journal of Physics: Conference Series, vol. 1258.
IOP Publishing, p. 012029.\\
Tanineah, D.M., Hussein, A.M., Widyan, H., Al-Wardat, M.A., 2023.
Trigonometric parallax discrepancies in space telescopes measurements
I: The case of the stellar binary system Hip 84976. Adv. Space Res. 71
(1), 1080–1088. https://doi.org/10.1016/j.asr.2022.09.025.\\
Tokovinin, A., 1997. Msc-a catalogue of physical multiple stars. Astron.
Astrophys. Suppl. Ser. 124 (1), 75–84.\\
Tokovinin, A., 2008. Comparative statistics and origin of triple and
quadruple stars., 389(2), 925–938. https://doi.org/10.1111/j.1365-2966.
2008.13613.x. arXiv:0806.3263.\\
Tokovinin, A., 2014a. From binaries to multiples. i. data on f and g dwarfs
within 67 pc of the sun. Astron. J. 147 (4), 86.\\
Tokovinin, A., 2014b. From binaries to multiples. ii. hierarchical
multiplicity of f and g dwarfs. Astron. J. 147 (4), 87.\\
Tokovinin, A., 2017. Orbit alignment in triple stars. Astrophys. J. 844 (2),
103. https://doi.org/10.3847/1538-4357/aa7746, arXiv:1706.00748.\\
Tokovinin, A., 2018. Spectroscopic orbits of subsystems in multiple stars.
iii. Astron. J. 156 (2), 48.\\
Tokovinin, A., 2020. Close binaries in hierarchical stellar systems.
Contrib. Astron. Obser. Skalnate Pleso 50 (2), 448–455. https://doi.
org/10.31577/caosp.2020.50.2.448.\\
Tokovinin, A., Mason, B.D., Hartkopf, W.I., Mendez, R.A., Horch, E.P.,
2016. Speckle Interferometry at SOAR in 2015. Astronom. J. 151 (6),
153. https://doi.org/10.3847/0004-6256/151/6/153, arXiv:1603.07596.\\
Tokovinin, A., Moe, M., 2020. Formation of close binaries by disc
fragmentation and migration, and its statistical modelling. 491(4),
5158–5171. https://doi.org/10.1093/mnras/stz3299. arXiv:1910.01522.\\
Toonen, S., Portegies Zwart, S., Hamers, A.S., Bandopadhyay, D., 2020.
The evolution of stellar triples. The most common evolutionary
pathways. Astronom. Astrophys. 640, A16. https://doi.org/10.1051/
0004-6361/201936835, arXiv:2004.07848.\\
Van Leeuwen, F., 1997. The hipparcos mission. Space Sci. Rev. 81 (3–4),
201–409.\\
Videla, M., Mendez, R.A., Claveria, R.M., Silva, J.F., Orchard, M.E.,
2022. Bayesian inference in single-line spectroscopic binaries with a
visual orbit. Astron. J. 163 (5), 220.\\
Vynatheya, P., Hamers, A.S., Mardling, R.A., Bellinger, E.P., 2022.
Algebraic and machine learning approach to hierarchical triple-star
stability. Mon. Not. R. Astron. Soc. 516 (3), 4146–4155.\\
Wenger, M., Ochsenbein, F., Egret, D., Dubois, P., Bonnarel, F., Borde,
S., Genova, F., Jasniewicz, G., Laloë, S., Lesteven, S., et al., 2000. The
simbad astronomical database-the cds reference database for astro-
nomical objects. Astron. Astrophys. Suppl. Ser. 143 (1), 9–22.

\end{document}